\documentclass{article}
\usepackage{arxiv}

\usepackage[utf8]{inputenc} 
\usepackage[T1]{fontenc}    
\usepackage{hyperref}       
\usepackage{url}            
\usepackage{booktabs}       
\usepackage{amsfonts}       
\usepackage{nicefrac}       
\usepackage{microtype}      
\usepackage{lipsum}		
\usepackage{graphicx}
\usepackage{tikz}
\usetikzlibrary{arrows,calc,positioning}
\usepackage{pgfplots}
\usepackage{graphicx}
\DeclareUnicodeCharacter{FB01}{fi}
\usepackage{caption}
\usepackage{csquotes}
\usepackage{colortbl}
\usepackage{multirow}
\usepackage{tabularx}
\usepackage{multicol}
\usepackage{rotating} 

\usetikzlibrary{arrows,shadows} 
\usepackage[underline=true,rounded corners=false]{pgf-umlsd}
\usepackage{enumitem}
\newlist{steps}{enumerate}{1}
\setlist[steps, 1]{label = Step \arabic*:}
\tikzstyle{intt}=[draw,text centered,minimum size=6em,text width=5.25cm,text height=0.34cm]
\tikzstyle{intl}=[draw,text centered,minimum size=2em,text width=2.75cm,text height=0.34cm]
\tikzstyle{int}=[draw,minimum size=2.5em,text centered,text width=3.5cm]
\tikzstyle{intg}=[draw,minimum size=.5em,text centered,text width=1.2cm]
\tikzstyle{intg1}=[draw,minimum size=1em,text centered,text width=1.8cm]
\tikzstyle{intg2}=[draw,minimum size=1em,text centered,text width=2.1cm]
\tikzset{%
   neuron missing/.style={
    draw=none, 
    scale=4,
    text height=0.333cm,
    execute at begin node=\color{black}$\vdots$
  },
}
\title{DATA SECURITY \& PRIVACY IN CLOUD COMPUTING: Concepts and Emerging Trends}
\author{ \href{https://orcid.org/0000-0001-9305-1269}{\includegraphics[scale=0.06]{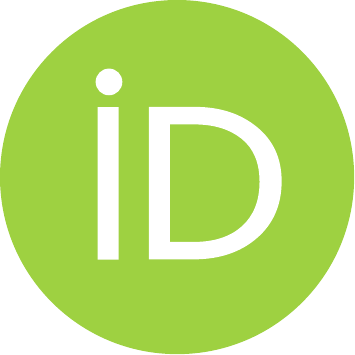}\hspace{1mm}Rishabh~Gupta}\thanks{This work is supported by the University Grant Commission, New Delhi, India, under the scheme of National Eligibility Test-Junior Research Fellowship (NET-JRF) with reference id-3515/(NET-NOV 2017)} \\
	Department of Computer Applications\\
	National Institute of Technology\\
	Kurukshetra, India \\
	\texttt{rishabh\_6180047@nitkkr.ac.in} \\
	\And
	\href{https://orcid.org/0000-0002-9689-6387}{\includegraphics[scale=0.06]{orcid.pdf}\hspace{1mm}Deepika Saxena}\thanks{The authors would like to thank National Institute of Technology Kurukshetra, India for financially supporting this research work.} \\
	Department of Computer Applications\\
	National Institute of Technology\\
	Kurukshetra, India \\
	\texttt{deepika\_6180096@nitkkr.ac.in} \\
	\And
	\href{https://orcid.org/0000-0002-8053-5050}{\includegraphics[scale=0.06]{orcid.pdf}\hspace{1mm}Ashutosh Kumar~Singh} \\
	Department of Computer Applications\\
	National Institute of Technology\\
	Kurukshetra, India \\
	\texttt{ashutosh@nitkkr.ac.in} \\
}

\renewcommand{\shorttitle}{Data Security \& Privacy in Cloud Computing}

\hypersetup{
pdftitle={A template for the arxiv style},
pdfsubject={q-bio.NC, q-bio.QM},
pdfauthor={Rishabh~Gupta, Ashutosh Kumar~Singh},
pdfkeywords={First keyword, Second keyword, More},
}
\begin{document}
\maketitle

\begin{abstract}
	Millions of users across the world leverages data processing and sharing benefits from cloud environment. Data security and privacy are inevitable requirement of cloud environment. Massive usage and sharing of data among users opens door to security loopholes. This paper envisages a disuscussion of cloud environment, its utilities, challenges, and emerging research trends confined to secure processing and sharing of data. 
\end{abstract}

\section{Introduction}
In 2006, San Jose proposed cloud computing's concepts in Search Engine Strategies (SES), and the National Institute of Standards and Technology (NIST) gave a formal definition of cloud. 
Almost every type of organization aspires to use cloud computing technology because of its features such as storage, computation, flexibility, scalability, etc \cite{saxena2021secure}. At present, data is developing at a faster pace. Indeed, even excessive information is being created every second, and likewise, the number of Internet-associated gadgets is increasing gradually \cite{kumar2020biphase}, \cite{singh2021quantum}. As the users may not have sufficient space to store their information, they need to keep their data on the cloud server and access the services facilitated by the cloud \cite{gupta2018probabilistic}, \cite{saxenaa2020communication}. Cloud storage gives an effect such that there is a vast space to store information and retrieve the data easily. Users have to pay for each utilized cloud service \cite{kumar2016dynamic}, \cite{saxena2020auto}. To make cloud computing possible and available to end-users, some services and models function behind the scenes. Fig. 1 shows two kinds of cloud computing models: deployment models and service models:
\tikzset{
  basic/.style  = {draw, text width=2cm, drop shadow, font=\sffamily, rectangle},
  root/.style   = {basic, rounded corners=2pt, thin, align=center,
                  fill=green!30},
  level 2/.style = {basic, rounded corners=6pt, thin,align=center, fill=green!60, align=center,
                  text width=8em, minimum size=.7cm},
  level 3/.style = {basic, thin, align=left, fill=pink!60,  align=center, text width=6em}
}
	\begin{figure}[ht!]
	\begin{center}
\begin{tikzpicture}
\begin{scope}[every node/.style={level 2}]
\node [xshift=-3cm,yshift=1cm](c1) {\small Cloud Charactertics};
\node [xshift=1cm,yshift=1cm] (c2){\small Service Models};
\node [xshift=5cm,yshift=1cm] (c3){\small Deployment Models};
\end{scope}
\begin{scope}[every node/.style={level 3}]
\node [below of = c1, xshift=15pt,yshift=-5pt] (c11) {\small On-demand self service};
\node [below of = c11,yshift=-7pt] (c12) {\small Broad network access};
\node [below of = c12,yshift=-7pt] (c13) {\small Resource pooling};
\node [below of = c13,yshift=-7pt] (c14) {\small Rapid elasticity or expansion};
\node [below of = c14,yshift=-35pt] (c15) {\small Measured service};
\node [below of = c2, xshift=15pt] (c21) {\small SaaS};
 \node [below of = c21, yshift=-1.3cm] (c22) {\small PaaS};
\node [below of = c22, yshift=-1.2cm] (c23) {\small IaaS};
\node [below of = c3, xshift=15pt] (c31) {\small Public};
\node [below of = c31] (c32) {\small Private};
\node [below of = c32] (c33) {\small Hybrid};
\node [below of = c33] (c34) {\small Community};
\end{scope}
\node [node distance=.1cm and -2.15cm,below left=of c21] (img1) {\includegraphics[width=0.1\textwidth,width=1.3cm]{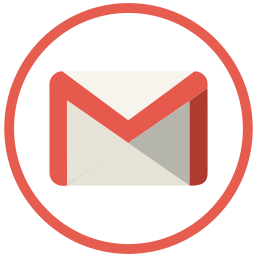}}; 
\node [node distance=.1cm and -2.1cm,below left=of c22] (img2) {\includegraphics[width=0.1\textwidth,width=1.3cm]{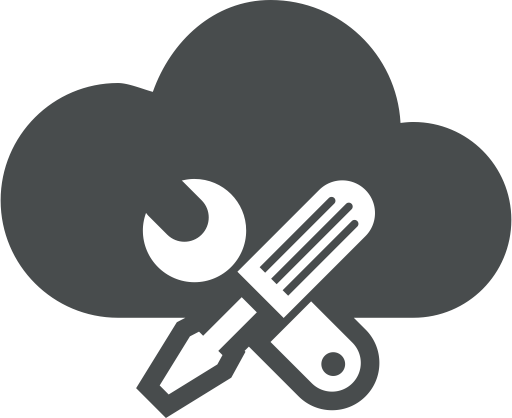}};
\node [node distance=.175cm and -2.2cm,below right=of c23] (img3) {\includegraphics[width=0.1\textwidth,width=1.3cm]{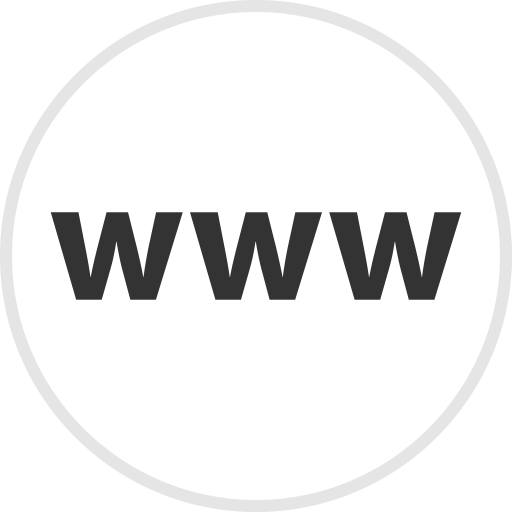}};
\foreach \value in {1,2,3,4,5}
  \draw[->] (c1.195) |- (c1\value.west);
\path (c14) -- (c15) node [black, font=\Huge, midway, sloped] {$\dots$};
\foreach \value in {1,...,3}
  \draw[->] (c2.195) |- (c2\value.west);
\foreach \value in {1,...,4}
  \draw[->] (c3.195) |- (c3\value.west);
\draw (1.5,-0.3) -- (1.5,-.55);
\draw (1.5,-2.6) -- (1.5,-2.84);
\draw (1.5,-4.8) -- (1.5,-5.085);
\end{tikzpicture}
	\end{center}
	\caption{Cloud Computing (NIST)}
	\label{fig:CloudComputing}
\end{figure}
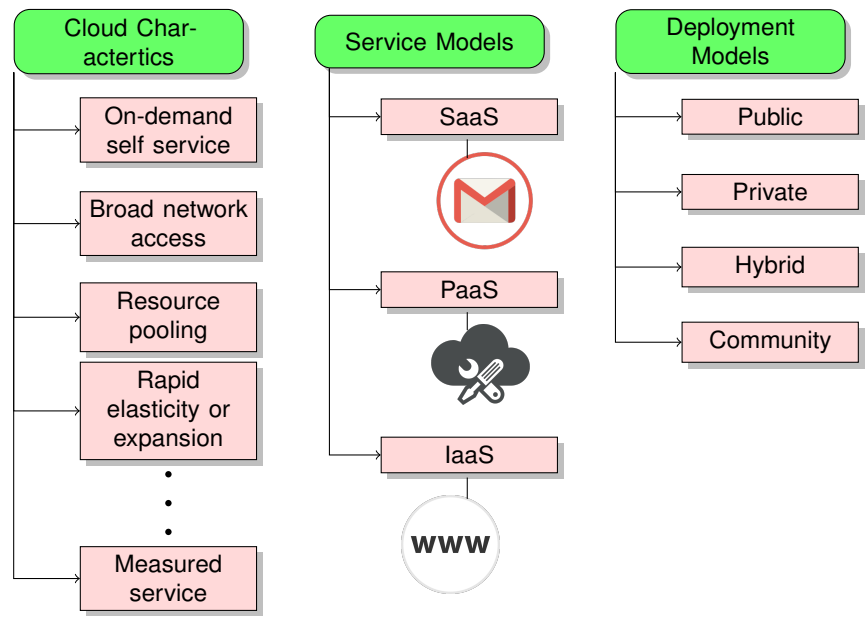
\par
The deployment models pertain to where the cloud infrastructure is located and managed. There are three commonly used cloud deployment models: public, private, and hybrid cloud. Another type of model is the community cloud which is commonly less used. A public cloud is a collection of computing resources that third-party organizations provide. It supports all users who want a computer resource, including subscription-based hardware (OS, CPU, memory, storage) or software (application server, database) to be used. A private cloud is typically one organization's infrastructure. The organization or a service provider can manage such infrastructure to help different customer groups. A hybrid cloud is a combination of private as well as public cloud computing resources. An additional model is a community cloud that shares in several organizations computing resources, and it is possible to manage either through organizational IT resources or third-party vendors \cite{saxena2020security}, \cite{saxena2021proactive}, \cite{saxena2021osc}. Cloud service models portray how customers can access the cloud \cite{gupta2019layer}. A combination of IaaS (infrastructure as a service), PaaS (platform as a service), and SaaS (software as a service) are the most basic service models. These service models can synergize; for instance, PaaS relies on IaaS because application systems require physical facilities \cite{singh2018data}.
\par
The model of IaaS offers customers parts of the infrastructure, including virtual machines, storage, networks, firewalls, and load balancers components \cite{chhabra2018probabilistic}. Customers have direct access to the lowest-level software like the virtual machine operating system, firewall, and load balancer dashboard in the stack with IaaS. Amazon Web Services is one of the largest IaaS providers. The PaaS model provides the customer with a pre-constructed application platform. Customers do not need to spend time building their applications ' underlying infrastructure. PaaS solutions usually offer an API that contains several features for the management and creation of programmatic platforms. Google AppEngine is one of the examples of a PaaS provider. SaaS offers prepared online software solutions. The SaaS software provider fully controls the application software. Online mail, project management systems and social media platforms are examples of SaaS applications \cite{singh2019sql}. 
\par
In 2012, Cisco Bonomi coined the term \enquote{fog computing}. The idea of fog computing is to bring the cloud closer to the devices. It is a distributed paradigm, and an extension of cloud computing. 
Fog computing's primary goal is to address the issues that cloud computing encounters during data processing. Fog is not a replacement for cloud computing; rather, these two technology complements each other. Routers, switches, intelligent devices, access points, and gateways are the essential parts of the fog layer.  These devices can store data, compute, route, and forward packets. The fog layer is the intermediate layer between user and cloud, as demonstrated in Fig. 2. This layer has rights of semi-permanent capacity that permit the impermanent information and data computation capability. The use of this layer has many advantages as low latency, higher real-time, broader geographical distributions. Moreover, it is observed that due to neighborhood information, the routing and storage costs in the fog environment are lesser than in the cloud environment \cite{wang2018three}. 
All of these benefits make fog computing more suitable for applications that are sensitive to delay.
	\begin{figure*}[!ht]
	\begin{center}
	\includegraphics[width=12cm]{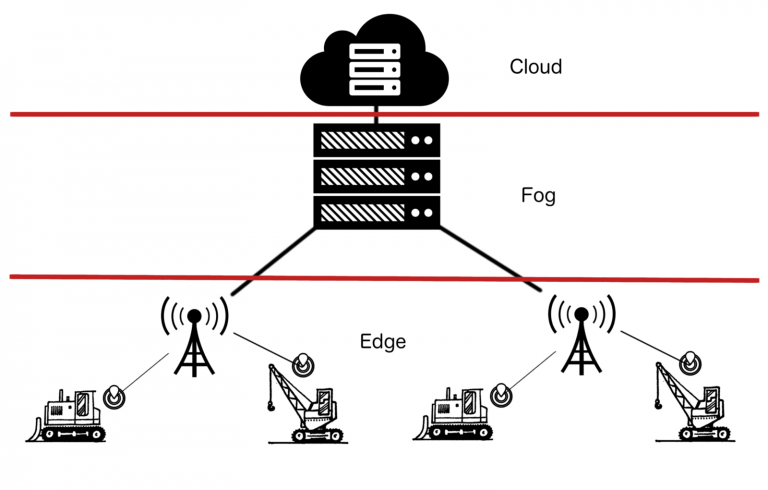}
		\caption{Fog Computing}
		\label{fig:StorageProcedure}
	\end{center}
\end{figure*}
\section{Motivation}
The paradigm of cloud computing has enabled various users to share resources \cite{saxena2021op}, \cite{singh2021cryptography}, \cite{kumar2021performance}. This notion has achieved broad popularity over the few years by constantly increasing the number of customers and supporting infrastructure development \cite{saxena2021workload}. For instance, Organizations like Cisco expect to have more than 24 billion gadgets connecting with the Internet by 2020. Further, Morgan Stanley has also predicted that almost 75 billion Internet devices will be in use by 2021. 
Cloud computing provides security to the customers, preventing their sensitive information and personal data from unauthorized stakeholders and multiple parties. Furthermore, data providers can monitor their outsourced data privacy at any moment. Organizations do not need to worry about data security because this facility is accessible as a service nowadays.
\par
Cloud computing is a growing technology that provides massive information without upfront investment to organizations with a novel business model \cite{kumar2020ensemble} \cite{chhabra2016dynamic}, \cite{singh2019secure}. However, most organizations are still reluctant to explore their business across the cloud because of security.  When the data is stored in the cloud server, the consequences are data ownership and management isolation \cite{chauhan2020survey}. Therefore, the Cloud service provider (CSP) can search and access data freely on the cloud server without taking the permission of the user. 
Meanwhile, the cloud server might be assaulted by an aggressor to get the user's information. There are many security issues in the cloud environment, such as man-in-the-middle attacks, data leaks, etc \cite{singh2020online}, \cite{deepika2020review}, \cite{chhabra2020secure}. All the above problems are exceptionally perils for user's information. Losses of the information of users is hampered, and there may occur data leakage problems too. If the user directly uploads data on the cloud, there may be issues like high bandwidth requirement, high latency, and a large volume of data. These are the biggest obstacles limiting cloud development. Therefore, it is necessary to work on it and find the right solution.
\par
Fig. 3 exhibits consumer fraud and identity theft complaints lodged during the period ranging from 2016 to 2020. It is observed that 4.8 million identity theft and fraud complaints were received in the year 2020. Out of these, 1.4 million complaints were identity theft, 25 percent of cases also reported money loss amounting, and remaining 2.2 million were fraud complaints. The number of complaints in 2020 is increased up to 46 percent over the previous year.
\begin{center}
\newcommand*{\captionsource}[2]{%
  \caption[{#1}]{%
    #1%
    \\\hspace{\linewidth}%
    \textbf{Source:} #2%
  }%
}
	\begin{figure*}[!h]
	\begin{center}
	\captionsetup{justification=centering}

	\includegraphics[width=14cm]{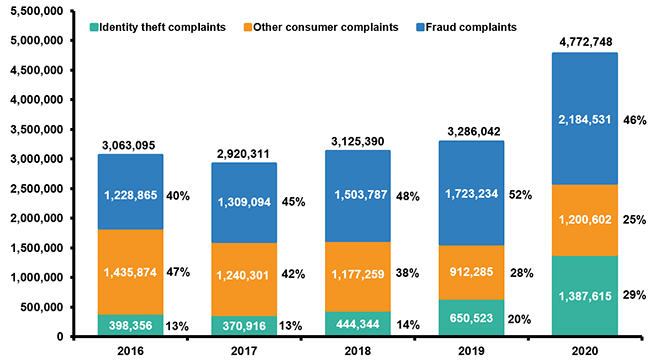}
 \centering
  \captionsource{Identity Theft And Fraud Reports, 2016-2020}{Federal Trade Commission, Consumer Sentinel Network}
	\end{center}
\end{figure*}
\end{center}
\par
To address the problems mentioned above, the state-of-art encryption schemes \cite{kaur2019digital}, \cite{kaur2017data} were used to protect the users' data. These schemes enable the users to encrypt their data and store it on the cloud servers. Afterward, cloud servers can carry out computations on it. But, Cloud servers are presently unable to provide innovative perspectives to users while fully maintaining their privacy. Therefore, efforts must be made to answer the question: \enquote{How can data consumers profit from cloud computing without compromising data privacy?}.
\section{Privacy-Preserving Based on Cryptography Mechanism}
Yuan and Yu \cite{yuan2013privacy} proposed a multiparty Back-Propagation Neural (BPN) network-based approach that is accurate, efficient, and secure for collaborative learning over arbitrarily partitioned data. To conduct operations over ciphertexts, they used a doubly homomorphic encryption technique. But they concentrated on enhancing data processing rather than the algorithm's efficiency.
Zhang et al. \cite{zhang2015privacy} proposed a privacy-preserving deep computation model based on homomorphic encryption. They used the offloading of the expensive operations to enhance the learning features of the cloud. The exponential process required by the Sigmoid function was utilized using the Taylor theorem. However, the model only includes addition and multiplication operations.
Yonetani et al. \cite{yonetani2017privacy} introduced a privacy-preserving mechanism based on a double-permitted homomorphic encryption (DPHE) scheme, which effectively learns visual classifiers across distributed private data. This scheme provided multiparty protected scalar products while minimizing the computational cost for high-dimensional classifiers. Nevertheless, either addition or multiplication operation can only support at a time.
A deep learning system based on additively homomorphic encryption was presented in \cite{aono2017privacy}. The proposed system protects gradients from the curious server. Asynchronous stochastic gradient descent (ASGD) trained overall participants' combined datasets obtained the same accuracy as the related deep learning system. However, the modified parameters are decrypted by the owner's secret key; thus, their model does not ensure parameter privacy.
A basic scheme based on multi-key fully homomorphic encryption (MK-FHE) mechanism was introduced in \cite{li2017multi}. The authors devised an advanced model for learning encrypted data in the cloud that uses the double decryption mechanism and fully homomorphic encryption (FHE) mechanism. But this scheme has a high cost in terms of computation and communication.
To perform the deep neural network algorithms over encrypted data, a framework named CryptoDL was proposed by Hesamifard et al. \cite{hesamifard2018privacy}. To address the existing limitations of homomorphic encryption schemes, they designed neural networking techniques. However, the proposed algorithm protects the owner's data by using keys that is not practical.
A privacy-preserving outsourced classification in cloud computing (POCC) framework was presented in \cite{li2018privacy}, which effectively enable an arbitrary number of multiplication and addition operations on ciphertexts. The data and query were protected by providing a proxy fully homomorphic encryption based on Gentry's scheme. Nevertheless, the cost of calculation and communication was increased in the proposed framework.
Ma et al. \cite{ma2018pdlm} proposed a privacy-preserving deep learning model, namely PDLM, to train the model over the data encrypted by the owners' keys. A privacy-preserving calculation toolset based on stochastic gradient descent (SGD) was utilized to accomplish the training task in a privacy-preserving way. Although the model minimized storage overhead, it has a high computation cost and lower classification accuracy.
A privacy-preserving outsourced classification scheme is presented in \cite{li2018outsourced}, which provides the classification services over encrypted data for users. They also designed two concrete secure classification protocols for the Naive Bayes classifier and the hyperplane decision-based classification, respectively. But during the launch of a classification query, user interactions are often involved in this scheme. 
Gao et al. \cite{gao2018privacy} proposed a privacy-preserving Naive Bayes classifier scheme that prevents information leakage under the substitution-then comparison (STC) attack. A double-blinding technique was adopted to protect the Naive Baye's privacy. Both the communication and processing overhead were decreased, but unable to discover the truth while maintaining privacy.
Phong and Phuong \cite{phuong2019privacy} proposed two systems, namely the Server-aided Network Topology (SNT) system and the Fully-connected Network Topology (FNT) system based on the connection with SNT and FNT server to protect the SGD privacy. The SNT and FNT systems achieved an accuracy corresponding to SGD using weight parameters instead of gradient parameters. These systems are both effective and efficient in terms of computing and communication.
Table 1 shows the summary of privacy-preserving of data based on the cryptography mechanism.
	\begin{table}[htbp]
		\caption{Summary of Privacy-Preserving based on Cryptography}
		\label{TableExp}
		\begin{center}
		\small\addtolength{\tabcolsep}{-4pt}
			\begin{tabular}{|c|c|c|c|}\hline 
				\textbf{\textit{Literature Reference}} & 
				\textbf{\textit{Approach}}&
					\textbf{\textit{Pros}} &  
						\textbf{\textit{Cons}} \\  \hline  \hline
 \multirow{1}{*}	{Wang et al. \cite{wang2018three}} & {Hash-Solomon code} & {High degree} & {High} \\ & {algorithm} & {privacy protection} & {Computation cost}   \\ \hline
\multirow{1}{*}  {Yuan et al. \cite{yuan2013privacy}  } & BGV fully and Doubly   & Encrypt data efficiently   & Low efficiency\\  & homomorphic encryption & secure scalar product & \\ \hline
\multirow{1}{*}  { Yonetani et al. \cite{yonetani2017privacy}} &  Doubly homomorphic  & Supported multi-party   & Both addition and \\ & encryption & secure scalar product & multiplication \\ \hline	
\multirow{1}{*}  {Li et al. \cite{li2017multi}} &  Multi-key fully  &  Preserve privacy of & Low efficiency  \\ &  homomorphic encryption & sensitive data & \\ \hline
\multirow{1}{*}  {Ma et al. \cite{ma2018pdlm}} &  Distributed Two Trapdoors & More  efficiency & Less accuracy  \\ &  Public-Key Cryptosystem  & & \\ \hline
\multirow{1}{*}	{Shokri and Shmatikov \cite{shokri2015privacy}} &  Optimization algorithms & Protect training data & High complexity \\ \hline
\multirow{1}{*}  {Wang et al. \cite{wang2017learning}} & Homomorphic encryption & Encrypted data  by randomly & Either addition or  \\ & &  splitting numerical &  multiplication \\ \hline
 \multirow{1}{*}	{Chan et al. \cite{chen2009privacy}} & {Homomorphic and } & {Partitioned Data} & {Two party} \\ & {ElGamal scheme} & {vertically} &    \\ \hline
 \multirow{1}{*}	{Bansal et al. \cite{bansal2011privacy}} & {Homomorphic and } & {Partitioned Data} & {Two party} \\ & {ElGamal scheme} & {arbitrarily} &    \\ \hline
  \multirow{1}{*}	{Samet et al. \cite{samet2012privacy}} & {BPNN and Extreme} & {Partitioned Data horizontally,} & {High Communication} \\ & {learning machine scheme} & { vertically with multi parties} &  {cost}  \\ \hline
 \multirow{1}{*}	{Cao et  al. \cite{cao2013privacy}} & {KNN,TF-IDF and} & {Solve the problem of muti-} & {Does not explore} \\ & {Dynamic Data Operation} & {Keyword ranked search} &  {checking integrity}  \\ \hline
 \multirow{1}{*}	{Guo et  al. \cite{guo2017dynamic}} & {TF-IDF model} & {Update the outsourced data} & {Does not improve} \\ & {} & {at less cost} &  {computational efficiency}  \\ \hline
 \multirow{1}{*}	{Fu et  al. \cite{fu2016toward}} & {Stemming algorithm } & {Outsourced Data With} & {Does not reflect } \\ & {Accuracy Improvement} & {at less cost} &  {the keyword weight}  \\ \hline
 \multirow{1}{*}	{Fu et  al. \cite{fu2018semantic}} & {Keyword based} & {Use two cloud server} & {High} \\ & {Search Scheme} & {for store and compute} &  {Compleaxity}  \\ \hline
\multirow{1}{*}	{Huang et  al. \cite{huang2017efficient}} & {Public key authenticated} & {Handle to resist} & {Single } \\ & {encryption} & {inside keyword} &  {keyword}  \\ \hline
\multirow{1}{*}	{Qi et  al. \cite{qi2017distributed}} & {Locality-Sensitive} & {Handle the service} & {Low} \\ & {Hashing} & {recommendation} &  {efficiency}  \\ \hline

			\end{tabular}
		\end{center}
	\end{table}
	
\section{Privacy-Preserving Based on Perturbation Mechanism}
Dwork et al. \cite{dwork2006calibrating} first proposed differential privacy and obtained complete background knowledge under the attacker's hypothesis. To protect data privacy, the randomly generated noise is disturbed according to a specially selected distribution.
Fletcher and Islam \cite{fletcher2017differentially} proposed a differential privacy decision-making random forest algorithm to reduce the query times and sensitivity. This scheme also minimizes the amount of noise that must be appended to protect the privacy and improve data availability. However, there is no consideration for the distributed scenario where multiple data owners conduct collaborative data mining.
To perform privacy-preserving machine learning over cloud data from different data providers, Li et al. \cite{li2018privacya} proposed a scheme that protects the data sets of various providers and the cloud. They used the public-key encryption with a double decryption algorithm (DD-PKE) to encrypt the data sets of the different providers with different public keys and $\epsilon$-differential privacy to add statistical noises into data to protect the privacy. Their scheme improved computational efficiency and data analysis accuracy. But such fully homomorphic (FHE) encryption schemes are commonly low efficiency.
A privacy-preserving Naive Bayes learning scheme with various data sources is presented in \cite{li2018differentially}. The proposed scheme enabled a trainer to train a Naive Bayes classifier over the dataset provided jointly by different data owners without the help of a trusted curator. However, collaboration is permitted, or adversaries can forge and modify the data in this scheme.
A distributed agent-based privacy-preserving framework, namely DADP, was proposed by Wang et al. \cite{wang2018privacy}. The proposed framework collects real-time spatial statistics data and publishes it with an untrusted server. To achieve global $w$-event $\epsilon$-differential privacy in a distributed manner, they utilized a distributed budget allocation mechanism and an agent-based dynamic grouping mechanism. The noise is added to crowd-sourced data using the Laplace technique in DAPM. It started a batch of reliable proxies (Agents) and anonymous connection technology to safeguard users' privacy under an untrusted server. Therefore, it was regarded as a semi-centralized setting and resulted in a more complex system.
An efficient privacy-preserving scheme based on machine learning was proposed by Hassan et al. \cite{hassan2019efficient}. Authors adopted a partially homomorphic encryption technique to encrypt data, and noised is added by applying a differential privacy mechanism. It allows all parties to publicly check the ciphertext's correctness via a low-cost unidirectional proxy re-encryption (UPRE) mechanism. However, the proposed system shared fewer data.
A private decision tree algorithm based on the noisy maximal vote was introduced in \cite{liu2018differentially}. To strike a balance between accurate counts and noise, an effective privacy budget allocation approach was utilized. The main aim of constructing an ensemble model is to increase the accuracy and stability by using differential privacy. The proposed algorithm performs the privacy analysis on each individual tree rather than the ensemble as a whole.
Gupta et al. \cite{gupta2020mlpam} proposed a machine learning and probabilistic analysis-based model, namely MLPAM. The authors used encryption, machine learning, and probabilistic approaches to share the multiple participants' data and minimize the risk affiliated with the leakage for prevention and detection. However, MLPAM is not able to give protection to the classifier.
To preserve the privacy of the data as well as query processing, Sharma et al. \cite{sharma2021differential} proposed a Differential Privacy Fuzzy Convolution Neural Network framework, namely DP-FCNN. The Laplace mechanism was utilized to inject the noise and encrypt the data by applying the lightweight Piccolo algorithm. The key properties were extracted using the BLAKE2s technique. But, DP-FCNN enhanced the computational overhead.
Table 2 summarises the privacy-preserving data using the Perturbation Mechanism.
	\begin{table}[htbp]
		\caption{Summary of Privacy-Preserving based on Perturbation}
		\label{TableExp}
		\begin{center}
		\small\addtolength{\tabcolsep}{-4pt}
			\begin{tabular}{|c|c|c|c|}\hline 
				\textbf{\textit{Literature Reference}} & 
				\textbf{\textit{Approach}}&
					\textbf{\textit{Pros}} &  
						\textbf{\textit{Cons}} \\  \hline  \hline
\multirow{1}{*}	{Fletcher and Islam \cite{fletcher2017differentially}} & {$\epsilon$-Differential Privacy} & {Reduce query time} & {Less accuracy}   \\ \hline
\multirow{1}{*}	{Li et  al. \cite{li2018privacya}} & {Homomorphic encryption} & {More efficiency} & {High Computation } \\ & {Differential Privacy} & {} &  {cost}  \\ \hline
\multirow{1}{*}	{Li et  al. \cite{li2018differentially}} & {$\epsilon$-Differential Privacy} & {Preserve privacy of data} & {Forge data } \\  \hline
\multirow{1}{*}	{Wang et al. \cite{wang2018privacy}} & {$w$ event} & {Data protection} & {Complex system} \\ & {$\epsilon$-Differential Privacy} & &  \\\hline
\multirow{1}{*}	{Hassan et al. \cite{hassan2019efficient}} & {Homomorphic encryption} & {Low cost} & {Limited data sharing} \\ & {$\epsilon$-Differential Privacy} & &  \\\hline
\multirow{1}{*}	{Liu et al. \cite{liu2018differentially}} & {Laplace mechanism} & {Balance between} & {Individual privacy} \\ &  & {accuracy and noise} &  \\\hline
\multirow{1}{*}	{Gupta et al. \cite{gupta2020mlpam}} & {Gaussian mechanism} & {High accuracy} & {No classifier} \\ &  &  & {protection} \\\hline
\multirow{1}{*}	{Sharma et al. \cite{sharma2021differential}} & {Laplace mechanism} & {Data protection} & {Computation overhead} \\ &  &  & {protection} \\\hline
			\end{tabular}
		\end{center}
	\end{table}
\section{Research Gaps}
On the basis of the literature review, the following research gaps are identified.
\begin{enumerate}
  \item Less protection, and privacy of the outsourced data. 
  \item   The efficiency of methods to protect privacy must be increased.
    \item Multiple user-based protection techniques are required.
    \item Computational and communication costs during the data transfer must be reduced.
    \item Minimization of the threats of data leakage during transmission.
\end{enumerate}
\section{Research Objectives}
To fill the identified study gaps, the goal described below is developed. 
\begin{enumerate}
\item To optimize the computation and communication costs among different entities.
 \item To reduce data leakage by using advanced encryption techniques.
\item To solve the security issue of cloud computing.
\item To formulate the approach to reduce the latency while preserving data privacy during transmission.
\end{enumerate}
\section{Justification for the objectives }
In an age of quickly growing information generation, it requires a computing infrastructure that can store and process enormous amounts of information. Developing these new cloud-based methods defines current safety and privacy demands and addresses consumer issues facing quality, effectiveness, and the best use of consumer needs. We provide complete awareness of secure cloud and efficient use of resources to validate our methodologies. Because security helps prevent data leakage and the disposal of data. With the following benefit, the advances to be carried out in research work will open an era for enhanced cloud facilities: 
\begin{enumerate}
    \item More quality of service (QoS)
    \item Better resource utilization
    \item Improves confidence in cryptography services 
    \item Low the cost of computation of the process
    \item Less service level agreement (SLA) violation
    \item Better use of the cloud for securely sharing of data among the organizations.
    \item Better efficiency
    \item Saving the cost
    \item Access the file universally
    \item Increase the security
\end{enumerate}
\bibliographystyle{IEEEtran}
\bibliography{references} 
\end{document}